\documentclass[12pt, a4paper]{article}

\usepackage{cite}
\usepackage{amsmath,amssymb}
\usepackage{comment}
\usepackage{multirow}
\usepackage[utf8]{inputenc}
\usepackage{bm}
\usepackage{accents}
\usepackage{cite}
\usepackage{braket}
\usepackage{url}

\usepackage{ifpdf}
\ifpdf
  \usepackage{graphicx, hyperref, xcolor}     
\else     
  \usepackage[dvipdfmx]{graphicx, hyperref, xcolor}     
 \fi
\usepackage{subcaption}

\definecolor{rossoferrari}{HTML}{D9073D}
\definecolor{mediumblue}{HTML}{0000CD}
\hypersetup{
setpagesize=false,
bookmarksnumbered=true,%
bookmarksopen=true,%
colorlinks=true,%
linkcolor=rossoferrari,
urlcolor=mediumblue,
citecolor=mediumblue,
}

\usepackage[height=21.5cm,width=16.5cm,centering]{geometry}

\begin{document}

\begin{titlepage}

\begin{center}

\hfill KEK-TH-2236

\hfill KEK-Cosmo-258

\vskip .75in

{\Large \bf
 Constraints on electron-scattering interpretation\\
 \vspace{3mm}
 of XENON1T excess
}

\vskip .75in

{\large
So Chigusa$^{(a)}$, Motoi Endo$^{(a,b,c)}$, and Kazunori Kohri$^{(a,b,c)}$
}

\vskip 0.25in

\begingroup\small\itshape
$^{(a)}$ \!\! KEK Theory Center, IPNS, KEK, Tsukuba, Ibaraki 305--0801, Japan
\\[0.3em]
$^{(b)}$ \!\! The Graduate University of Advanced Studies (Sokendai), Tsukuba, Ibaraki 305--0801, Japan
\\[0.3em]
$^{(c)}$ \!\! Kavli Institute for the Physics and Mathematics of the
  Universe (WPI), University of Tokyo,
  Kashiwa 277-8583, Japan
\endgroup

\end{center}
\vskip .5in

\begin{abstract}
Recently, the XENON1T experiment has observed an excess in the electronic recoil data in the recoil energy range of $1$--$7$\,{\rm keV}.
One of the most favored new physics interpretations is electron scattering with a boosted particle with a velocity of $\sim 0.1$ and a mass of $\gtrsim 0.1\,\mathrm{MeV}$.
If such a particle has a strong interaction with electrons, it may affect the standard scenario of cosmology or be observed at low-threshold direct detection experiments.
We study various constraints, mainly focusing on those from the big-bang nucleosynthesis, supernova cooling, and direct detection experiments.
We discuss the implication of these constraints on electron-scattering interpretation of the XENON1T excess.
\end{abstract}

\end{titlepage}


\renewcommand{\thepage}{\arabic{page}}
\setcounter{page}{1}
\renewcommand{\thefootnote}{\#\arabic{footnote}}
\setcounter{footnote}{0}

\section{Introduction}

The XENON1T experiment has recently reported an excess in low-energy electronic recoil data~\cite{Aprile:2020tmw}.
The observed number of events in the recoil energy range of 1--7\,{\rm keV} is 285 which is larger than the known background expectation $232 \pm 15$, corresponding to the significance of $> 3\sigma$.
Although the excess might be due to unknown background contributions, e.g., from $\beta$-decay of tritium, it could be a sign of new physics~\cite{Smirnov:2020zwf, Takahashi:2020bpq, Kannike:2020agf, Alonso-Alvarez:2020cdv, Amaral:2020tga, Fornal:2020npv, Boehm:2020ltd, Bally:2020yid, Harigaya:2020ckz, Su:2020zny, Du:2020ybt, DiLuzio:2020jjp, Chen:2020gcl, Dey:2020sai, Bell:2020bes, Buch:2020mrg, AristizabalSierra:2020edu, Choi:2020udy, Paz:2020pbc, Lee:2020wmh, Robinson:2020gfu, Cao:2020bwd, Khan:2020vaf, Primulando:2020rdk, Nakayama:2020ikz, Jho:2020sku, Bramante:2020zos, Baryakhtar:2020rwy, An:2020bxd, Zu:2020idx, Gao:2020wer, Budnik:2020nwz, Lindner:2020kko, Chala:2020pbn, Bloch:2020uzh, Dent:2020jhf, DeRocco:2020xdt, Zioutas:2020cul, McKeen:2020vpf, Coloma:2020voz, An:2020tcg, DelleRose:2020pbh, Dessert:2020vxy, Bhattacherjee:2020qmv, Ge:2020jfn, Chao:2020yro, Gao:2020wfr, Ko:2020gdg, Cacciapaglia:2020kbf, Alhazmi:2020fju, Sun:2020iim, Baek:2020owl, Szydagis:2020isq, Croon:2020ehi, Li:2020naa}.
One of the most attractive scenarios is provided by the dark matter (DM).
If the DM has a mass of $\sim 1$\,keV, and is absorbed by electrons in the detector, the XENON1T result can be explained.
However, the event distribution of the electronic recoil energy has a sharp peak around the DM mass, though the XENON1T result implies a broader spectrum for the excess~\cite{Aprile:2020tmw}.

Such a broad spectrum may be realized by a new particle $\chi$ scattering with electrons~\cite{Kannike:2020agf, Fornal:2020npv, Su:2020zny, Jho:2020sku, Ko:2020gdg, DelleRose:2020pbh, Alhazmi:2020fju, Baek:2020owl}.
It is likely to be electrically neutral because events with multiple scatterings are vetoed in the analysis~\cite{Aprile:2020tmw}.\footnote{
The particle is allowed to have a magnetic moment.
}
The recoil energy of $\sim 1$\,keV can typically be realized by $\chi$ with a velocity of $\sim 0.1$ with a mass of $\gtrsim 0.1$\,MeV.\footnote{
The excess can also be explained by the boosted DM scattering with $v \sim 1$ and the mass $\ll 0.1$\,MeV~\cite{Alhazmi:2020fju}. 
The following constraints are applicable to such a case.
}
The energy deposition becomes smaller than the keV scale when it is lighter or if the velocity is lower, e.g., as low as the virial velocity of the cold DM, $v \sim 10^{-3}$.

The velocity is so large that $\chi$ cannot be trapped in gravitational potentials in the universe. 
Hence, its energy abundance must be suppressed in the early universe.
In this paper, $\chi$ is assumed to acquire such a large velocity by some boost mechanism.
Such a situation has been discussed in Refs.~\cite{Huang:2013xfa, Agashe:2014yua, Detmold:2014qqa, Kim:2016zjx, Kamada:2017gfc}.
For example, in the semi-annihilation DM model~\cite{DEramo:2010keq, Kamada:2017gfc, Belanger:2012zr, Belanger:2014bga, Hektor:2019ote},\footnote{
See, e.g., Refs.~\cite{Kamada:2018hte, Kamada:2019wjo} for studies on phenomenology.
} the DM is boosted via the annihilation process of $\chi \chi \to \chi X$ with $m_X \ll m_\chi$, which takes place in the Galactic Center or halo.\footnote{
It is argued in Ref.~\cite{Fornal:2020npv} that the flux of $\chi$ becomes insufficient to explain the XENON1T result if the annihilation takes place in the Sun.
}
Then, $\chi$ in the final state has a boost factor of $\gamma = (5m_\chi^2 - m_X^2)/4m_\chi^2$.
Depending on the scenario when the boost happens, we will explore the following two cases:
\begin{enumerate}

\item Models with a mechanism that boosts $\chi$ in the {\it current} universe.
It is noticed that $\chi$ is not always a substantial component of the DM.
It is identical with the DM in the semi-annihilation DM model, while it can be different if the boost takes place via $X X \to \chi_1 \chi_2$ with $m_{\chi_{1,2}} < m_X$, where $X$ is the DM, and the stable (or long-lived) $\chi_1$ or $\chi_2$ is detected.
The relic abundance of $\chi_1$ and $\chi_2$ depends on the thermal history.

\item Models in which $\chi$ was boosted in the {\it early} universe.
Here, $\chi$ is required to be stable because it is detected at XENON1T.
Since $\chi$ is relativistic for the redshift $z \gtrsim 10$, it must not be responsible for the DM.
In this paper, we do not specify its production mechanism.

\end{enumerate}
In order to realize the XENON1T excess, $\chi$ is assumed to interact with electrons effectively.
Then, it may affect the standard $\Lambda$CDM cosmology or be detected by direct detection experiments apart from the latest XENON1T result~\cite{Aprile:2020tmw}.
In this paper, we will examine the following constraints:\footnote{
We will consider the mass range of $>0.1$\,MeV.
For a smaller mass region, see Ref.~\cite{Lehmann:2020lcv}.
}
\begin{itemize}

\item If the mass of $\chi$ is smaller than $\sim 10$\,MeV, $\chi$ can be produced thermally by electron-positron annihilations during the Big-Bang nucleosynthesis (BBN).
Its abundance contributes to the expansion rate of the universe, affecting the primordial abundances of the light nuclei, especially D/H.

\item For the mass $\lesssim 100$\,MeV, $\chi$ is generated from the thermal plasma via the electron-positron annihilation inside a core-collapse supernova (SN).
Then, $\chi$ escapes from the SN with carrying the energy or is trapped inside the SN by losing its kinetic energy via the scattering with the SN materials, i.e., the electrons.
Thus, the emission of $\chi$ enhances the energy loss rate of the SN when its free-streaming length is large enough, while it is trapped and does not affect the SN cooling process if the interaction with the electron is too large.
Thus, the annihilation rate $e^+e^- \to \chi\chi$ is constrained from two sides.

\item In the first model, $\chi$ can be a substantial component of the DM.
Then, the un-boosted component of $\chi$ may also be detected by low-threshold direct detection experiments of the DM, e.g., the XENON1T S2-only analysis~\cite{Aprile:2019xxb}, even though the electronic recoil energy is low.
The experiments can constrain the $e$-$\chi$ scattering cross section for the mass $\gtrsim 10$\,MeV.

\end{itemize}
Besides, the scattering cross section of $\chi$ with electrons may be constrained by the observation of the structure formation of the universe and the anisotropy of the cosmic microwave background (CMB) as long as $\chi$ is abundant in the early universe.
In the first model, if $\chi$ is a substantial component of the DM and is in kinetic equilibrium with the thermal plasma, the density fluctuation of the DM is washed out by the $e$-$\chi$ scattering.
Also, when $\chi$ annihilates into the electron and positron, the thermal plasma receives extra energy and the effective number of relativistic degrees freedom, $N_{\mathrm{eff}}$, is modified. 
In addition, if $\chi$ is relativistic in the early universe as supposed in the second model, its abundance contributes to $N_{\mathrm{eff}}$.
These contributions can affect thermal history.
We will study the implication of these constraints on electron-scattering interpretation of the XENON1T excess.

This paper is organized as follows.
We describe the model setup in Sec.~\ref{sec-setup}.
The constraints are summarized in Sec.~\ref{sec-constraint}, where the bound on the cross section is provided.
Based on the XENON1T result, the constraint on the cross section is interpreted to the limit on the flux/abundance of $\chi$.
In Sec.~\ref{sec-result}, the results are shown for the two models.
Section~\ref{sec-conclusion} is devoted to conclusion.

\section{Model}
\label{sec-setup}

In this paper, $\chi$ represents a particle which is a source of the XENON1T excess, but we do not necessarily identify $\chi$ as the DM.
We consider $\chi$ to be an electrically neutral Dirac fermion, which is stable (or long-lived) and interacts with an electron $e$ via a contact interaction term,\footnote{
The following analysis is independent of the chirality of electron.
If the left handed lepton is involved in the interaction term, the $\chi$ annihilation into a pair of neutrinos also puts constraints on the model \cite{Arguelles:2019ouk}.
However, since the constraints are likely to be weaker than those from interaction with electron, we only focus on the later in this paper.
}
\begin{align}
  \mathcal{L}_{\mathrm{int}} = G_{\chi e} (\bar{\chi}\gamma_\mu\chi) (\bar{e}\gamma^\mu e).
  \label{eq-L-int}
\end{align}
This setup corresponds to the case when the interaction is mediated by a heavy vector field whose mass exceeds the center-of-mass energy of the collisions we will consider below.\footnote{
It is straightforward to apply the following analysis to other types of $e$-$\chi$ interactions.
In particular, if the mediator is light, such a particle can also contribute to the constraints from BBN, SN cooling and the structure formation, depending on its mass (see, e.g., Ref.~\cite{Knapen:2017xzo}).
}
In the XENON1T detector, we assume $\chi$ has a velocity $v_\chi \sim 0.1$ to deposit recoil energy in 1--7\,keV via the scattering process of $e^{-}\chi \to e^{-}\chi$~\cite{Kannike:2020agf}. 
In the non-relativistic limit, the corresponding cross section $\sigma_0$ is obtained as
\begin{align}
  \sigma_0 = \frac{G_{\chi e}^2 \mu_{\chi e}^2}{\pi},
\end{align}
where $\mu_e$ is the induced mass,
\begin{align}
  \mu_{\chi e} = \frac{m_\chi m_e}{m_\chi+m_e},
\end{align}
with $m_\chi$ and $m_e$ being masses of $\chi$ and $e$, respectively.
Note that we are interested in the mass range of $m_\chi \gtrsim 0.1$\,MeV to explain the XENON1T result~\cite{Kannike:2020agf}. 
In addition, $\chi$ affects the BBN and SN cooling via the annihilation process $e^+ e^- \to \bar{\chi} \chi$. 
Its cross section is expressed in terms of $\sigma_0$ as 
\begin{align}
  \sigma_{\mathrm{ann}} (s) =
  \begin{cases}
    \dfrac{\sigma_0}{12\mu_{\chi e}^2}
    \dfrac{\sqrt{s-4m_\chi^2}}{\sqrt{s-4m_e^2}}
    \dfrac{(s+2m_e^2) (s+2m_\chi^2)}{s}, & (s>4m_\chi^2)\\
    0, & (s< 4m_\chi^2)
  \end{cases}
  \label{eq-sigma-ann}
\end{align}
where $s$ is the Mandelstam variable. 
In this paper, we do not include other interactions of $\chi$ with the SM particles for simplicity. 
They are likely to strengthen the following constraints. 

\section{Constraints}
\label{sec-constraint}

In this section, we discuss various constraints on $\chi$ arising from the contact interaction \eqref{eq-L-int}.
We will explain the constraint from BBN in Sec.~\ref{sec-BBN}, that from SN cooling in Sec.~\ref{sec-SN}, and that from direct detection experiments in Sec.~\ref{sec-DD}, all of which give important constraints on the mass and cross section of $\chi$.
We will also comment on the constraint from structure formation in Sec.~\ref{sec-structure} and that from $N_{\mathrm{eff}}$ in Sec.~\ref{sec-neff}.

\subsection{Big-Bang nucleosynthesis}
\label{sec-BBN}

If $\chi$ is in thermal equilibrium with the baryon-photon plasma at the beginning of BBN with temperature $T_{\mathrm{BBN}} \sim {\cal O}(1)\,\mathrm{MeV}$, thermally produced $\chi$ particles affect the expansion rate of the universe.
Such unusually high expansion rate modifies the primordial abundances of the light nuclei and is severely constrained, in particular from the observation of the D/H abundance~\cite{Kolb:1986nf, Serpico:2004nm}.

Instead of solving Boltzmann equations, we adopt the criteria that the thermally averaged electron-positron annihilation rate $\Gamma_{e^+e^-\to\bar{\chi}\chi} \equiv n_e \Braket{\sigma v}_{\mathrm{ann}}$ is smaller than the Hubble parameter $H$ for the BBN to work successfully.\footnote{
We neglect the effects of the late-time entropy generation and the photo-dissociation due to the residual $\chi$ annihilation during the freeze-out epoch because it is argued in Ref.~\cite{Serpico:2004nm} that their contributions are subdominant.  
}
Here, $n_e \sim T^3$ denotes the electron number density with $T$ being temperature of the thermal plasma. 
The thermally averaged annihilation cross section is calculated as~\cite{Gondolo:1990dk}
\begin{align}
  \Braket{\sigma v}_{\mathrm{ann}} = \frac{1}{N} \int_{4m_e^2}^\infty ds\,
  \sigma_{\mathrm{ann}}(s) \sqrt{s} (s-4m_e^2) K_1 \left( \frac{\sqrt{s}}{T} \right),
\end{align}
with
\begin{align}
  N = 8m_e^4 T K_2^2 \left( \frac{m_e}{T} \right).
\end{align}
Here, $K_n$ $(n=1,2,\dots)$ is the modified Bessel function of the second kind.
Using these quantities, the condition is given by
\begin{align}
  \left. \frac{\Gamma_{e^+e^-\to\bar{\chi}\chi}}{H} \right|_{T=T_{\mathrm{BBN}}} > 1.
  \label{eq-BBN}
\end{align}
The Hubble parameter is estimated as $H(T)\sim \sqrt{g_{*}(T)} T^2/M_{\mathrm{pl}}$, where $g_{*}$ is the effective number of relativistic degrees of freedom, which takes the value of $g_{*}\simeq 10.75$ at $T=T_{\mathrm{BBN}}$, and $M_{\mathrm{pl}} \simeq 1.22\times 10^{19}\,\mathrm{GeV}$ is the Planck mass.
Consequently, the condition is approximately obtained as
\begin{align}
  \sigma_0 \lesssim
  \begin{cases}
    4\times 10^{-43}\,\mathrm{cm}^2 \left( \dfrac{m_\chi}{\mathrm{MeV}} \right)^2, & (m_\chi < m_e)\\
    10^{-43}\,\mathrm{cm}^2. & (m_e < m_\chi \lesssim 10\,\mathrm{MeV})
  \end{cases}
\end{align}
This result is valid when $\chi$ is highly relativistic. 
In the numerical analysis, we calculate the above condition without using this approximation for the cross section \eqref{eq-sigma-ann}. 
Note that the constraint is significantly weaker if $m_\chi \gtrsim 10\,\mathrm{MeV}$, which is natural because such heavy particles are barely produced in the thermal bath at $T=T_{\mathrm{BBN}} \sim {\cal O}(1)\,\mathrm{MeV}$.
The result is consistent with those in previous analyses such as Ref.~\cite{Sabti:2019mhn}.

\begin{figure}[t]
  \centering
  \includegraphics[width=0.6\hsize]{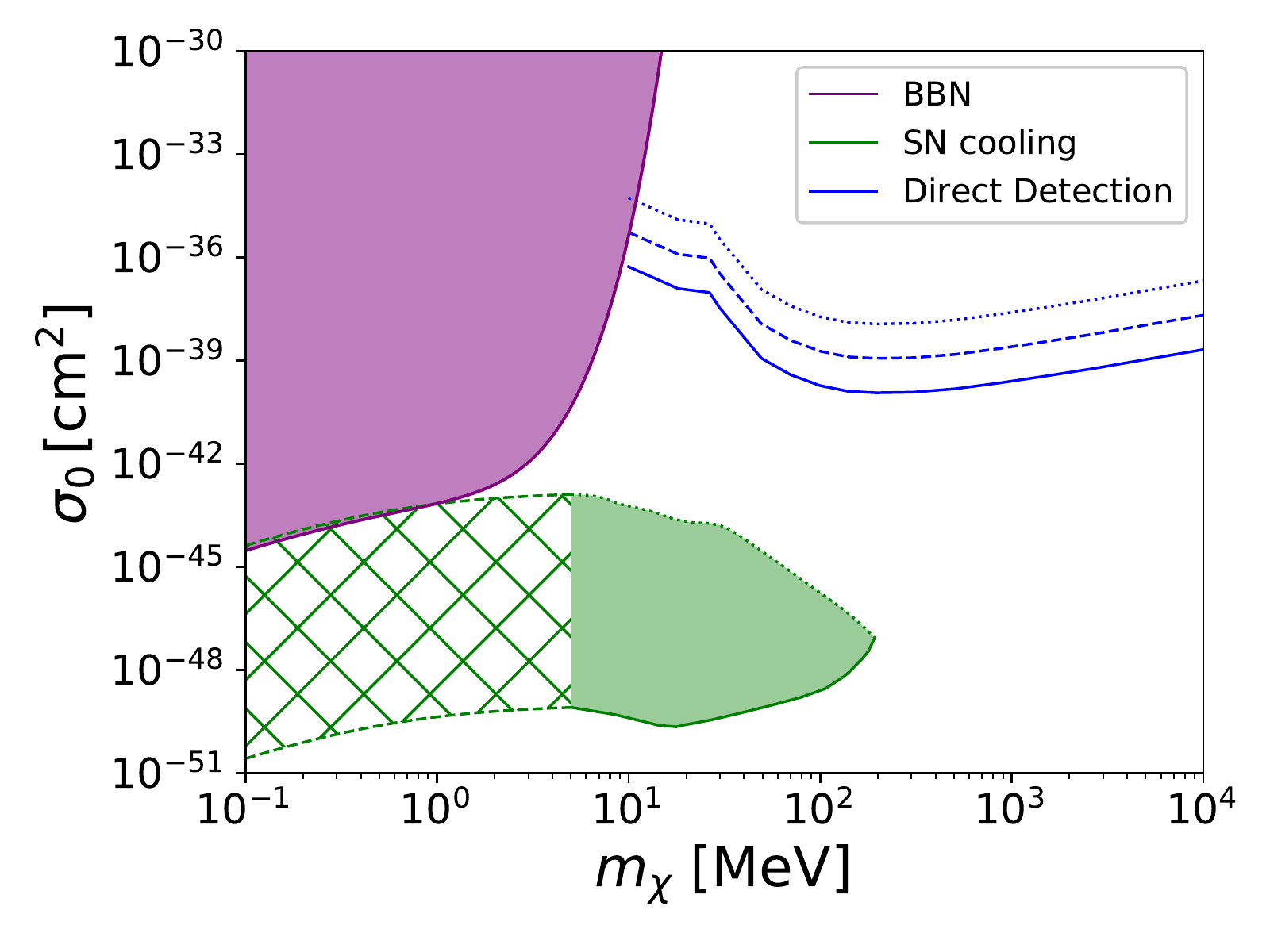}
  \caption{
    Constraints on the $e$-$\chi$ interaction $G_{\chi e}$ in terms of $\sigma_0$ as a function of $m_\chi$. 
    The purple and green shaded (or hatched) regions are disfavored by BBN~\cite{Sabti:2019mhn} and SN cooling~\cite{DeRocco:2019jti}, respectively.
    The regions above blue solid, dashed, and dotted lines are also excluded by direct detection experiments~\cite{Essig:2017kqs, Aprile:2019xxb,Agnes:2018oej,Agnes:2018ves,Agnes:2018fwg} if $\chi$ consists of $100\%$, $10\%$, $1\%$ of the DM energy density in the current universe, respectively.
  }
  \label{fig-constraint-sigma}
\end{figure}

In Fig.~\ref{fig-constraint-sigma}, we show the constraints on the coupling strength $G_{\chi e}$ in terms of $\sigma_0$ as a function of $m_\chi$.
The purple shaded region is excluded by the BBN constraint.
We can see that the cross section is severely constrained from above for $m_\chi \lesssim 10\,\mathrm{MeV}$.

It is noticed that the interaction rate becomes larger when the temperature becomes higher in our setup \eqref{eq-L-int}, as it is clearly shown from the $s$-dependence of the cross section in Eq.~\eqref{eq-sigma-ann}.
Thus, it may be possible that $\chi$ and electron are in thermal equilibrium at $T=T_R$, with $T_R$ being the reheating temperature, while the equilibrium is no longer maintained at $T=T_{\mathrm{BBN}}$.
If this is the case, $\chi$ particles are thermally created at $T=T_R$ and the relic of $\chi$ may affect BBN at $T=T_{\mathrm{BBN}}$.\footnote{
In this case, the relic of $\chi$ at $T=T_{\mathrm{BBN}}$ highly depends on the decoupling temperature that is determined by the interaction strength.
In general, the contribution to $N_{\mathrm{eff}}$ is likely to be suppressed if the decoupling occurs before QCD phase transition.
}
However, since the situation highly depends on the reheating temperature $T_R$ in this case, here we adopt a model-independent and conservative bound evaluated at $T=T_{\mathrm{BBN}}$ as in Eq.~\eqref{eq-BBN} with the observational lower bound on $T_R$ to be $T_R \gtrsim {\cal O} (1)$~MeV ~\cite{Kawasaki:1999na,Kawasaki:2000en,Ichikawa:2005vw,deSalas:2015glj,Hasegawa:2019jsa}.

\subsection{Supernova cooling}
\label{sec-SN}

If there exist extra light degrees of freedom that couple with the thermal plasma inside stars such as SN, they are generated and emitted from the star and contribute to the stellar cooling.
By comparing the standard prediction of the cooling with its observation, we can put a bound on the interaction strength of the light degrees of freedom (for example, see Ref.~\cite{Mohapatra:1990vq}).

The SN cooling constraint on pair production of a Dirac fermion $\chi$ has been discussed in detail based on Monte Carlo simulation in Ref.~\cite{DeRocco:2019jti}, where $\chi$ is assumed to couple with the SM electromagnetic current $j_\mu^{\rm em}$ via a contact interaction $\bar\chi\gamma^\mu\chi j_\mu^{\rm em}$.
Although both the electron-positron annihilation $e^+ e^- \to \bar{\chi} \chi$ and the proton-neutron bremsstrahlung $pn \to pn\bar\chi\chi$ primarily contribute to the production of $\chi$, it was mentioned that the former process dominates for $r\gtrsim 5$\,km in all $\chi$ mass region inside a SN, and thus, governs the cooling process.
In this paper, we convert the constraint on the interaction strength shown in Fig.~6 of Ref.~\cite{DeRocco:2019jti} to that on $G_{\chi e}$ by assuming that the boundary is determined by the electron processes. 

However, this conversion may not be accurate in the trapped regime.
If the interaction strength is large enough, the produced $\chi$ loses its kinetic energy via the scattering with the SN materials and is trapped inside the SN eventually, leaving the SN cooling unaffected.
In contrast to the model setup in Ref.~\cite{DeRocco:2019jti}, $p$-$\chi$ interactions are absent in this paper.
Therefore, the trapping is less efficient and the converted limit we obtain is considered to be conservative in the trapped regime.

Besides, since the constraint in Ref.~\cite{DeRocco:2019jti} was displayed only for $m_\chi > 5\,\mathrm{MeV}$, we need to extrapolate it down to $m_\chi = 0.1\,\mathrm{MeV}$.
This could be done by an observation that the constraint on the interaction strength becomes less sensitive to $m_\chi$ in the region $m_\chi \ll T_C \sim \mathcal{O}(10)\,\mathrm{MeV}$ with $T_C$ being the core temperature of SN.
In fact, the results provided in Ref.~\cite{Guha:2018mli} show such behavior, though their analysis was not based on the Monte Carlo simulation.
Hence, we extrapolate the result in Ref.~\cite{DeRocco:2019jti} by assuming that the constraint on the interaction strength is independent of $m_\chi$ in $m_\chi < 5\,\mathrm{MeV}$.

The result is shown by the green-colored region in Fig.~\ref{fig-constraint-sigma}.
In particular, the constraint obtained by the extrapolation is shown by the hatched area with dashed boundaries.
From the figure, we can see that a wide range of $0.1\,\mathrm{MeV} \lesssim m_\chi \lesssim 100\,\mathrm{MeV}$ is constrained by the SN cooling.
The constraint is placed on the region with smaller cross sections than those from BBN.
In the region below the green area, the production rate of $\chi$ is tiny enough to keep the cooling unaffected, while the region above is also allowed because the produced $\chi$ is trapped inside the SN.
According to the above argument, the upper boundary, which is denoted by the green dotted line, is conservative.
We expect that the SN cooling may put a constraint in a larger cross-section region. 
Nonetheless, to obtain more precise constraints, we need to perform a detailed Monte Carlo simulation, which is beyond the scope of this paper.

\subsection{Direct detection experiments}
\label{sec-DD}

If $\chi$ is a substantial component of the DM relic abundance in the current universe, its scattering cross section is also constrained by the low-threshold direct detection experiments searching for electron recoils such as XENON10/100/1T experiments~\cite{Essig:2017kqs,Aprile:2019xxb}.
Before boosted, given the DM velocity of $\sim 10^{-3}$, the electronic recoil energy is less than $\sim 1$\,keV. 
Such low energy has been searched for especially at the XENON1T experiment; the S2-only analysis has a sensitivity on the electronic recoil energy of $\sim 100$\,eV~\cite{Aprile:2019xxb}.

The $e$-$\chi$ scattering cross section is severely constrained in the sub-GeV to GeV mass range of our interest.
For $m_\chi \gtrsim 30$\,MeV, the recent XENON1T result~\cite{Aprile:2019xxb} provides the best sensitivity, while the cross section is constrained by XENON10 below it~\cite{Essig:2017kqs}.
The results in the mass range $m_\chi = 100\,\mathrm{MeV}$--$1\,\mathrm{GeV}$ are summarized in Fig.~5 of Ref.~\cite{Aprile:2019xxb}.
In Fig.~\ref{fig-constraint-sigma}, we show the constraint by the blue line.
Here, the solid, dashed, and dotted lines correspond to the case when $\chi$ constitutes $100\%$, $10\%$, and $1\%$ of the total amount of the DM energy density, respectively.
For each line, the region above the line is excluded.

Let us comment on the constraints from the indirect searches. 
If $\chi$ is a substantial component of the DM and has a large annihilation cross section for $\bar\chi\chi \to e^+e^-$, we may observe anomalous electron and positron cosmic rays.
Although interstellar sub-GeV electrons/positrons are shielded by the solar magnetic field, Reference~\cite{Boudaud:2016mos} has provided a constraint on the annihilation cross section in this range by utilizing the Voyager 1 result as well as the AMS-02 one.
The result is $\langle \sigma_{\rm ann}v \rangle \lesssim 5 \times 10^{-29}\,\mathrm{cm^3 s^{-1}}$ for $m_\chi \sim 100$\,MeV and the limit is weaker for the other masses, where it is assumed that $\chi$ accounts for the total amount of the DM relic abundance.
Thus, the constraint is much weaker than the one from the direct detection in Fig.~\ref{fig-constraint-sigma}.

\subsection{Structure formation}
\label{sec-structure}

If $\chi$ occupies a substantial component of the DM and is in kinetic equilibrium with the baryon-photon plasma at some time in the cosmological history, the density fluctuation of the DM which seeds the structures of the universe is washed out via the acoustic oscillation of the plasma.
The matter power spectrum is suppressed under the scale which enters the horizon before the time~\cite{Sigurdson:2003vy, Profumo:2004qt}.
Expressing the scale of the fluctuation as $k$, the horizon crossing occurs when $k=da/dt$ with $a$ being the scale factor.
Then, $t=t_s\sim 10^6\,\mathrm{s}$ or $T=T_s\sim 1\,\mathrm{keV}$ corresponds to the damping scale $k^{-1}\sim 10^2\,\mathrm{kpc}$~\cite{Kohri:2009mi}, which is the order of the galaxy scale.
We require that $\chi$ is not in the kinetic equilibrium with the thermal plasma at $T=T_s$ for the successful formation of galaxies.\footnote
{
If the kinetic equilibrium is maintained at the time of recombination with $T=T_{\mathrm{rec}}\simeq 0.26\,\mathrm{eV}$, it can also distort the CMB anisotropy~\cite{McDermott:2010pa} if $\chi$ occupies more than $6\%$ of the abundance of the DM~\cite{Dubovsky:2003yn}.
However, it can be read off from Eq.~\eqref{eq-time-scale} that the higher temperature results in a much larger interaction rate.
Accordingly, the constraint from galaxy formation is much severer than that from CMB anisotropy because $T_s > T_{\mathrm{rec}}$, and thus we only focus on the former.
}
This criterion regarding the kinetic equilibrium is expressed as
\begin{align}
  t_\chi (T_s) > t_s,
  \label{eq-structure}
\end{align}
where $t_\chi(T)$ denotes the time scale for $\chi$ to achieve the kinetic equilibrium with SM particles at temperature $T$, which will be evaluated below.
This condition should be satisfied as long as $\chi$ is abundant in the universe.

The time scale $t_\chi(T)$ is estimated as follows (cf.,~Ref.~\cite{McDermott:2010pa}).
Let us consider the non-relativistic limit of the $e$-$\chi$ scattering in the center-of-mass frame.\footnote
{
Note that the non-relativistic approximation used here is valid when $m_\chi, m_e \gg T$.
The temperature $T_s$ of our interest satisfies this condition.
}
After a collision, the $\chi$ momentum changes by an amount of $\delta p_\chi$,
\begin{align}
  \delta p_\chi^2 = 2\mu_{\chi e}^2 v_{\mathrm{rel}}^2 (1-\cos\theta_{*}),
\end{align}
where $v_{\mathrm{rel}}$ denotes the relative velocity, and $\theta_{*}$ is the scattering angle.
Then, the thermally averaged momentum transfer per unit time is evaluated as
\begin{align}
  \frac{d\Braket{\delta p_\chi^2}}{dt} =
  n_e \int d^3v_e d^3v_\chi d\Omega_{*}\, f(v_e)f(v_\chi) \frac{d\sigma_0}{d\Omega_{*}} v_{\mathrm{rel}}\, \delta p_\chi^2,
\end{align}
where $\Omega_{*}$ is the solid angle in the center-of-mass frame, $f(v)$ is the Maxwell-Boltzmann distribution function, and $d\sigma_0/d\Omega_{*}$ is the differential scattering cross section.
On the other hand, the thermally averaged momentum squared of $\chi$ in its comoving frame is
\begin{align}
  \Braket{p_\chi^2} =  \int d^3v_\chi\, f(v_\chi) (m_\chi v_\chi)^2 = 3m_\chi T.
\end{align}
By combining the quantities defined above, the time scale $t_\chi$ is obtained as
\begin{align}
  t_\chi(T) = \frac{\Braket{p_\chi^2}}{d\Braket{\delta p_\chi^2}/dt}.
\end{align}
Since the $e$-$\chi$ scattering is isotropic in the non-relativistic limit, $d\sigma_0/d\Omega_{*} = \sigma_0/4\pi$ is satisfied.
By substituting this into the above expressions, we obtain
\begin{align}
  t_\chi (T) \simeq \frac{3\sqrt{\pi} m_\chi}{8 n_e \mu_{\chi e}^{1/2} T^{1/2} \sigma_0},
  \label{eq-time-scale}
\end{align}
in our model.
Recalling Eq.~\eqref{eq-structure}, this gives an constraint on the cross section
\begin{align}
  \sigma_0 \lesssim \begin{cases}
    10^{-28}\,\mathrm{cm}^2 \left( \dfrac{m_\chi}{\mathrm{MeV}} \right)^{1/2}, & (m_\chi < m_e) \vspace{1mm} \\
    10^{-28}\,\mathrm{cm}^2 \left( \dfrac{m_\chi}{\mathrm{MeV}} \right), & (m_\chi > m_e)
  \end{cases}
\end{align}
where we use $n_e(T_s) \simeq n_e(T_{\mathrm{rec}}) (T_s/T_{\mathrm{rec}})^3$, with $n_e(T_{\mathrm{rec}})\simeq 2.58\times 10^{-12}\,\mathrm{eV}^3$~\cite{WMAP_params} being the electron number density at the recombination with temperature $T_{\mathrm{rec}}\simeq 0.26\,\mathrm{eV}$.
This constraint is, however, much looser than the others in Fig.~\ref{fig-constraint-sigma}. 

\subsection{$N_{\mathrm{eff}}$}
\label{sec-neff}

The CMB observations precisely measure the effective number of relativistic degrees freedom, $N_{\mathrm{eff}}$, defined by
\begin{align}
  N_{\mathrm{eff}} \equiv \frac{8}{7} \left( \frac{11}{4} \right)^{4/3}
  \left( \frac{\rho_{\mathrm{rad}} - \rho_\gamma}{\rho_\gamma} \right),
  \label{eq-Neff}
\end{align}
where $\rho_{\mathrm{rad}}$ and $\rho_\gamma$ are the energy density of radiation and photons, respectively.
Within the standard model, $N_{\mathrm{eff}}$ counts the number of neutrino species under the massless approximation and we obtain $N_{\mathrm{eff}}\sim 3$ at the current universe.
The CMB observations conclude $N_{\mathrm{eff}} \simeq 2.99\pm 0.17$~\cite{Aghanim:2018eyx}, which is consistent with the standard model.
Thus, if there exist new light degrees of freedom after the neutrino decoupling at $T=T_D\sim 2\,\mathrm{MeV}$~\cite{Dolgov:2002wy} and before the recombination at $T=T_{\mathrm{rec}}$, it may modify $N_{\mathrm{eff}}$.
Since the contributions to $N_{\mathrm{eff}}$ depend crucially on whether $\chi$ is non-relativistic or relativistic in $T_{\mathrm{rec}} < T < T_D$, we will discuss them separately.

Let us first discuss the effect of the non-relativistic $\chi$. 
For instance, this corresponds to the case when $\chi$ is a component of the DM.
In this case, the contribution to $N_{\mathrm{eff}}$ is only through the $\chi$ annihilation into electrons.
Since the energy exchange between electrons and photons is still active at $T=T_D$, the energy deposit from $\chi$ effectively heats the baryon-photon plasma and increases its temperature.
As a result, $N_{\mathrm{eff}}$ of our interest, which is proportional to the ratio between the energy densities of neutrino and photon as shown in Eq.~\eqref{eq-Neff}, decreases.
Although the effect of the $e$-$\chi$ interaction on $N_{\mathrm{eff}}$ has been evaluated in Refs.~\cite{Henning:2012rm, Slatyer:2015jla, Sabti:2019mhn}, 
the constraint was shown to be weaker than that from BBN.

On the other hand, there are two possible effects on $N_{\mathrm{eff}}$ when $\chi$ is relativistic in $T_{\mathrm{rec}} < T < T_D$.
One is the same as that explained above; $\chi$ deposits energy on the baryon-photon plasma and $N_{\mathrm{eff}}$ decreases.
The other is the contribution from $\chi$ itself to $\rho_{\mathrm{rad}}$ in Eq.~\eqref{eq-Neff}, resulting in an increase of $N_{\mathrm{eff}}$.
For evaluating their contributions quantitatively, we need a detailed numerical analysis of Boltzmann equations, and we leave it as a future work.

\section{Implication on interpretation of XENON1T}
\label{sec-result}

The recent result of the XENON1T experiment~\cite{Aprile:2020tmw} suggests the existence of a particle with velocity $v_\chi \sim 0.1$ and mass $m_\chi \gtrsim 0.1\,\mathrm{MeV}$ that necessarily interacts with electrons.
Such an interpretation requires a large flux or a large interaction strength to explain the signal rate at XENON1T.
In the previous section, we studied the constraints on the interaction strength.
Hence, a bound on the $\chi$ flux is derived by requiring enough signal rate. 
In the following, we will discuss these constraints in the two models listed in the Introduction; one with $\chi$ boosted in the current universe and the other with $\chi$ boosted in the early universe.

\subsection{Models with $\chi$ boosted in the current universe}

To be explicit, let us discuss the boosted DM model~\cite{Huang:2013xfa, Agashe:2014yua, Detmold:2014qqa, Kim:2016zjx, Kamada:2017gfc}.
In particular, the particle $\chi$ is considered to be boosted via the semi-annihilation of DMs~\cite{DEramo:2010keq, Kamada:2017gfc, Belanger:2012zr, Belanger:2014bga, Hektor:2019ote}.\footnote
{
The following analysis can be applied to models with other boost mechanisms. 
See, e.g., Refs.~\cite{SungCheon:2008ts, Belanger:2011ww} for multi-component DM models.
} 
According to Ref.~\cite{Fornal:2020npv}, the flux $\Phi_\chi$ of $\chi$ should be as large as
\begin{align}
  \Phi_\chi \sim 5\times 10^2\,\mathrm{cm}^{-2}\,\mathrm{s}^{-1}
  \left( \frac{10^{-36}\,\mathrm{cm}^2}{\sigma_0} \right),
  \label{eq-flux-sigma}
\end{align}
where $\sigma_0$ is the $e$-$\chi$ scattering cross section, and the total number of $\chi$ events at XENON1T is fixed to be $\sim 50$~\cite{Aprile:2020tmw}.
The size of $\Phi_\chi$ depends on the boost mechanism. 
If the semi-annihilation $\chi\chi \to \chi X$ with $m_X \ll m_\chi$ takes place in the Galactic Center or halo, it is estimated as (cf.,~Ref.~\cite{Agashe:2014yua})\footnote{
In Ref.~\cite{Agashe:2014yua}, two boosted DMs are generated by an annihilation process.
Correspondingly, we divide their expression by two to derive Eq.~\eqref{eq-BDM-flux}, which denotes the flux for semi-annihilation associated with only one boosted DM.
}
\begin{align}
  \Phi_{B} \sim 10^2\,\mathrm{cm}^{-2}\,\mathrm{s}^{-1}
  \left( \frac{\Braket{\sigma v}_{\mathrm{DM}}}{5\times 10^{-26}\,\mathrm{cm}^3\,\mathrm{s}^{-1}} \right)
  \left( \frac{\mathrm{MeV}}{m_\chi} \right)^2,
  \label{eq-BDM-flux}
\end{align}
where $\Braket{\sigma v}_{\mathrm{DM}}$ denotes the thermal average of the DM annihilation cross section in the current universe, and the NFW profile of the DM distribution~\cite{Navarro:1995iw} is assumed.

\begin{figure}[t]
  \centering
  \includegraphics[width=0.6\hsize]{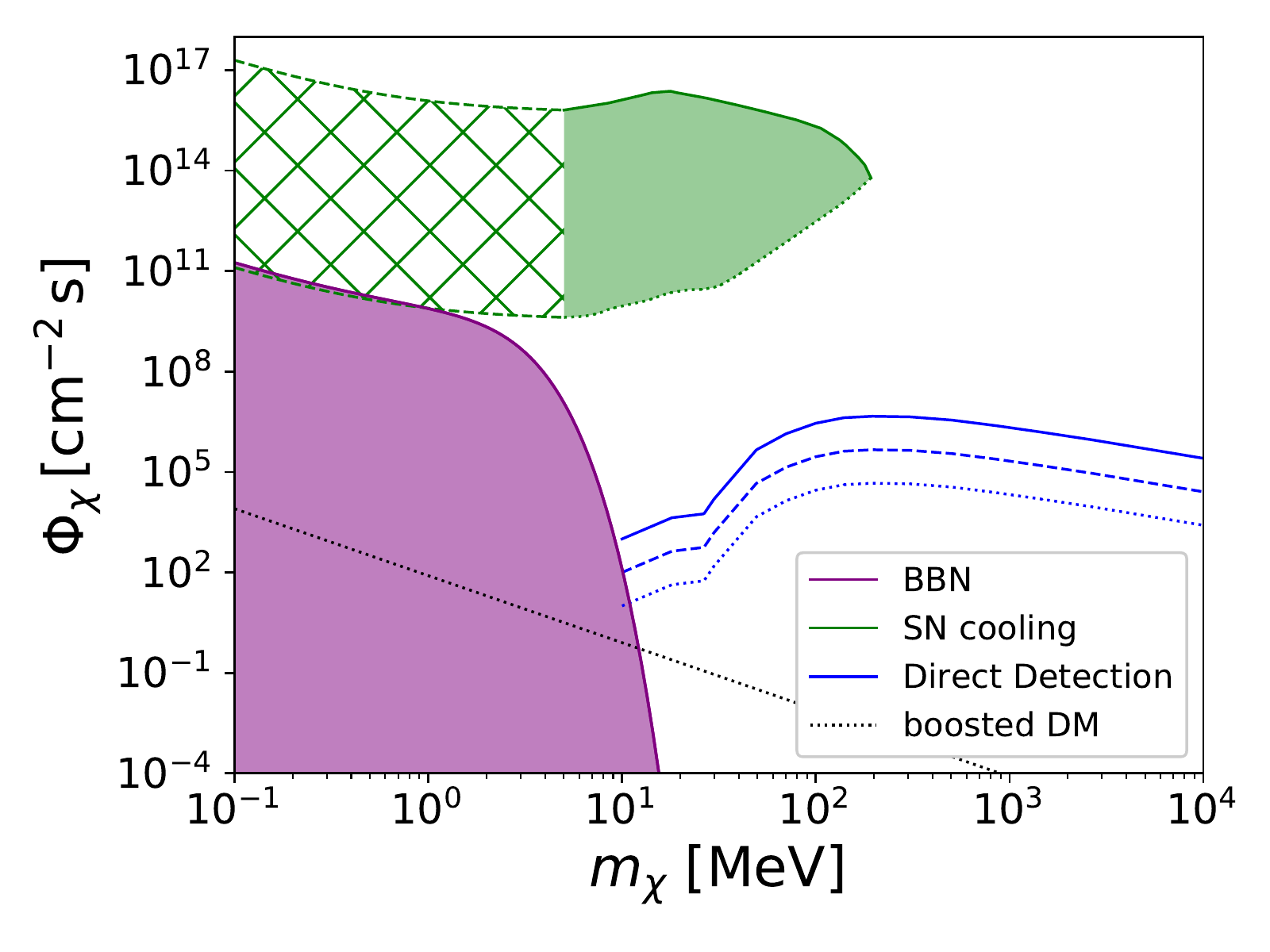}
  \caption{
    Constraints on the flux of $\chi$ that explains the XENON1T excess.
    The color and style convention is the same as Fig.~\ref{fig-constraint-sigma}.
    The black dotted line denotes the expected flux of $\chi$ in the boosted DM model with $\Braket{\sigma v}_{\mathrm{DM}} = 5\times 10^{-26}\,\mathrm{cm}^3\,\mathrm{s}^{-1}$.
  }
  \label{fig-constraint-flux}
\end{figure}

By using Eq.~\eqref{eq-flux-sigma}, the flux of $\chi$ is related to its scattering cross section which is constrained in the last section.
In Fig.~\ref{fig-constraint-flux}, we show the constraints on $\Phi_\chi$ as a function of $m_\chi$.
The purple, green, and blue colors correspond to the constraints from BBN, SN cooling, and direct detection experiments, respectively.
As mentioned in Sec.~\ref{sec-constraint}, the blue solid, dashed, and dotted lines correspond to the cases when $\chi$ consists of $100\%$, $10\%$, and $1\%$ of the total amount of the DM energy density, respectively.
Also shown in the black dotted line is the expected values of the $\chi$ flux evaluated in Eq.~\eqref{eq-BDM-flux}. 
Here, we set $\Braket{\sigma v}_{\mathrm{DM}} = 5\times 10^{-26}\,\mathrm{cm}^3\,\mathrm{s}^{-1}$ as a reference, which corresponds to the standard value for the weakly interacting particle regime.

For $m_\chi \lesssim 10\,\mathrm{MeV}$, the flux is required to be enhanced by many orders of magnitude larger than the black dotted line in order to explain the XENON1T excess without confronting the SN bound.
Such a large flux is beyond the naive expectation, and we need some new mechanism to enlarge it, e.g., a cascade shower of light elements from heavy DM decays.
For $m_\chi \gtrsim 10\,\mathrm{MeV}$, the severe constraint is obtained from the direct detection experiments if $\chi$ constitutes a sizable component of the DM, as in the model of semi-annihilation.\footnote{
The relic abundance of $\chi$ depends on models.
See, e.g., Ref.~\cite{Belanger:2011ww} for a study of two-component DM models. 
}
Therefore, this model, i.e., the one that $\chi$ is boosted in the current universe, works if there is a mechanism to generate a huge flux, or if $\chi$, which is observed at XENON1T, does not consist of the DM.
In the latter case, although some of the parameter space is still constrained by SN cooling, there remain large allowed regions.
Note that the allowed region that is compatible with the naively expected flux shown by the black dotted line corresponds to the trapping regime for $m_\chi \lesssim 100\,\mathrm{MeV}$, where the interaction strength is too strong to contribute to the SN cooling.

We comment on other possible constraints that are not displayed in the figure.
As discussed in Sec.~\ref{sec-structure}, there is a bound from the structure formation.
It is either weaker than the BBN constraint ($m_\chi \lesssim 10\,\mathrm{MeV}$) or too weak to be shown in the figure ($m_\chi \gtrsim 10\,\mathrm{MeV}$).
Also, there is a bound from $N_{\mathrm{eff}}$, which is always weaker than that from BBN as discussed in Sec.~\ref{sec-neff}.
Note that these two constraints are applied if there is a sufficiently large abundance of $\chi$ at $T=T_s$ and $T_{\mathrm{BBN}}$, respectively.

\subsection{Models with $\chi$ boosted in the early universe}

On the contrary to the idea explored so far, it is also possible that there exist particles $\chi$ with high velocity $v_\chi\sim 0.1$ that are originally generated in the early universe.
Let $\rho_{\mathrm{DM}}$ ($\rho_\chi$) be the energy density of the DM (the high velocity particle $\chi$).
We also define the fraction $f_\chi$ of $\chi$ over the current DM abundance $f_\chi \equiv \rho_\chi / \rho_{\mathrm{DM}}$.
Even though $\chi$ is not a part of the DM, $f_\chi < 1$ is required to avoid the overclosure of the universe.
Assuming that the boosted particle $\chi$ explains the XENON1T excess, Fig.~3 of Ref.~\cite{Kannike:2020agf} tells us the best fit value,
\begin{align}
  n_\chi \sigma_0 \sim 4\times 10^{-44}\,\mathrm{cm}^{-1}
  \label{eq-Kannike}
\end{align}
where $n_\chi = \rho_\chi / m_\chi$ is the number density of $\chi$.
Here, $n_\chi$ is related to $f_\chi$ as
\begin{align}
  n_\chi = f_\chi \frac{\rho_{\mathrm{DM}}}{m_\chi} \sim 300\,\mathrm{cm}^{-3}\, f_\chi
  \left( \frac{\mathrm{MeV}}{m_\chi} \right),
\end{align}
where the energy density of the DM around the earth $\rho_{\mathrm{DM}} \sim 0.3\,\mathrm{GeV}/\mathrm{cm}^3$~\cite{Kerr:1986hz,Green:2011bv} is used.
Thus, for a fixed value of $m_\chi$, larger $\sigma_0$ is required to maintain Eq.~\eqref{eq-Kannike} when $f_\chi$ decrease; this correlation results in the constraints on $f_\chi$ from the constraint in Sec.~\ref{sec-constraint}.

\begin{figure}[t]
  \centering
  \includegraphics[width=0.6\hsize]{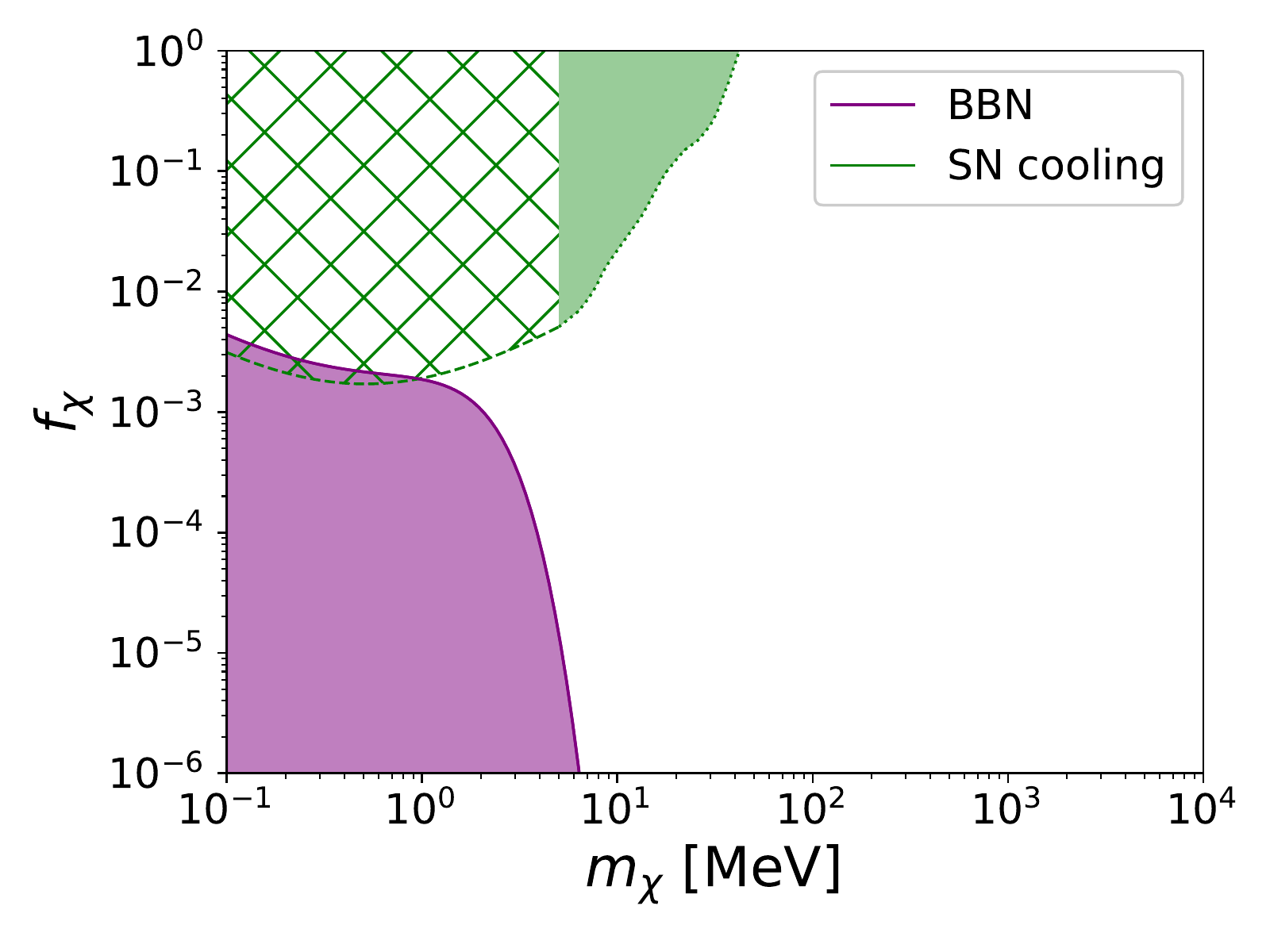}
  \caption{
    Constraints on the fraction of $\chi$, $f_\chi=\rho_\chi / \rho_{\mathrm{DM}}$, as a function of $m_\chi$.
    The color and style convention is the same as Fig.~\ref{fig-constraint-sigma}.
  }
  \label{fig-constraint-fraction}
\end{figure}

In Fig.~\ref{fig-constraint-fraction}, we show the constraint on $f_\chi$ as a function of $m_\chi$.
The purple and green colors denote the constraints from BBN and SN cooling, respectively.
Again, the shaded (or hatched) regions are excluded by the corresponding observation.
For $m_\chi \lesssim 10\,\mathrm{MeV}$, our model is severely constrained by BBN and SN cooling, and there is no allowed region with $f_\chi \leq 1$.
For $m_\chi \gtrsim 10\,\mathrm{MeV}$, a large parameter space remains unconstrained.
We need to be aware, however, that a large $f_\chi$ may result in a large abundance of relativistic freedom in the early universe considering the redshift of the energy, which may conflict with the small scale structure of the DM halo.
According to the arguments on the warm DM, there is a constraint on the fraction of the relativistic freedom $f_{\mathrm{rel}} \lesssim 0.1$ that contributes to the structure formation~\cite{Viel:2005qj}.
Since $f_\chi$ is the fraction in the current universe and $f_{\mathrm{rel}}$ is that in the early universe, the relationship between $f_\chi$ and $f_{\mathrm{rel}}$ highly depends on the generation mechanism of $\chi$.
It is likely that some model building regarding the generation of $\chi$ is needed to realize $f_\chi \gtrsim \mathcal{O}(0.1)$, though we do not discuss in detail here.

We comment on further model-dependent constraints that are not displayed in the figure.
If $\chi$ particles exist at the time of recombination, they affect the observed spectrum of CMB.
Considering the redshift of the energy, the currently non-relativistic $\chi$ with velocity $v_\chi\sim 0.1$ is anticipated to have relativistic velocity $v_\chi\sim 1$ at large redshift $z\gtrsim 10$, e.g., at the recombination.
Then, the effect of this relativistic $\chi$ is, as described in Sec.~\ref{sec-neff}, to change $N_{\mathrm{eff}}$ through the energy deposit on the baryon-photon plasma and the contribution to the radiation energy density.
Since this constraint highly depends on the generation mechanism of $\chi$, we need model-dependent treatment for the quantitative analysis and refrain from analyzing here.

\section{Conclusion}
\label{sec-conclusion}

In this paper, we have considered cosmological, astrophysical, and experimental constraints on a new particle $\chi$ that has a contract interaction with electron \eqref{eq-L-int}.
It turned out that the strong interaction with electrons leads to various constraints mainly from BBN, SN cooling, and the direct detection experiments.
We have summarized the obtained constraint on the mass $m_\chi$ and the scattering cross section $\sigma_0$ in Fig.~\ref{fig-constraint-sigma}.
A strong upper bound on $\sigma_0$ is provided by BBN for $m_\chi \lesssim 10\,\mathrm{MeV}$, while the SN cooling constrains a region with smaller $\sigma_0$ for $m_\chi \lesssim 100\,\mathrm{MeV}$.
Also, the mass range of $m_\chi=10\,\mathrm{MeV}$--$10\,\mathrm{GeV}$ is searched for sensitively by the XENON experiments if $\chi$ constitutes a sizable part of the DM.

Next, we have discussed the implication of the above constraints on the electron-scattering interpretation of the XENON1T excess.
We have focused on two models: one with a boost mechanism of $\chi$ in the current universe, and the other with boosted $\chi$ that is generated in the early universe.
For the first model, we have shown constraints in Fig.~\ref{fig-constraint-flux} in terms of the flux of the currently boosted $\chi$.
It was concluded that this model works if there is a mechanism to generate a huge $\chi$ flux, or if $\chi$ does not constitute the DM.
For the second model, we have shown constraints in Fig.~\ref{fig-constraint-fraction} in terms of the fraction of the boosted $\chi$ over the DM relic abundance in the current universe.
In this model, we have found that models with $m_\chi\lesssim 10\,\mathrm{MeV}$ are severely constrained by BBN and SN cooling, while there remains a large parameter space allowed for $m_\chi \gtrsim 10\,\mathrm{MeV}$.

\section*{Acknowledgments}

This work is supported in part by JSPS KAKENHI Grant ([SC] and No.~JP17H01131 [KK]),
Early-Career Scientists (No.~16K17681 [ME]), 
MEXT KAKENHI Grant (19H05114 [KK], 20H04750 [KK]), and World Premier Interna-
tional Research Center Initiative (WPI Initiative), MEXT, Japan (ME and KK).

\bibliography{xenon}
\bibliographystyle{JHEP}

\end{document}